\documentclass[aps,prl,10pt,twocolumn,superscriptaddress,balancelastpage,showpacs,reprint,longbibliography]{revtex4-1}
\usepackage[english]{babel}
\usepackage[utf8]{inputenc}
\usepackage[colorinlistoftodos, color=green!40, prependcaption]{todonotes}

\usepackage{centernot}
\usepackage{graphicx}
\usepackage{amsmath}
\usepackage{times}
\usepackage{amssymb}
\usepackage{mathrsfs}
\usepackage{color}
\usepackage{url}
\usepackage{version}
\usepackage[pdftex,colorlinks=true,pdfstartview=FitV,linkcolor= linkcolor,citecolor= linkcolor,urlcolor= linkcolor,hyperindex=true,hyperfigures=false]{hyperref}
\usepackage{balance}
\usepackage{lastpage}
    
\definecolor{linkcolor}{rgb}{0,0,0.6} 

\newcommand{\qqq}{\end{eqnarray}}

\newcommand{\bfr}{\mathbf{r}}
\newcommand{\bfp}{\mathbf{p}}

\newcommand{\bfU}{\mathbf{U}}

\begin{document}

\title{Symmetric mixtures of pusher and puller microswimmers behave as noninteracting suspensions}

\author{D\'ora B\'ardfalvy}
\affiliation{Division of Physical Chemistry, Lund University, P.O. Box 124, S-221 00 Lund, Sweden}

\author{Shan Anjum}
\affiliation{Division of Physical Chemistry, Lund University, P.O. Box 124, S-221 00 Lund, Sweden}

\author{Cesare Nardini}
\affiliation{Service de Physique de l'\'Etat Condens\'e, CNRS UMR 3680, CEA-Saclay, 91191 Gif-sur-Yvette, France}
\affiliation{Sorbonne Universit\'e, CNRS, Laboratoire de Physique Th\'eorique de la 
Mati\`ere Condens\'ee, LPTMC, F-75005 Paris, France.}

\author{Alexander Morozov}
\affiliation{SUPA, School of Physics and Astronomy, The University of Edinburgh, James Clerk Maxwell Building, Peter Guthrie Tait Road, Edinburgh, EH9 3FD, United Kingdom}

\author{Joakim Stenhammar}
\email{joakim.stenhammar@fkem1.lu.se}
\affiliation{Division of Physical Chemistry, Lund University, P.O. Box 124, S-221 00 Lund, Sweden}

\date{\today}

\begin{abstract}
Suspensions of rear- and front-actuated microswimmers immersed in a fluid, known respectively as ``pushers'' and ``pullers'', display qualitatively different collective behaviours: beyond a characteristic density, pusher suspensions exhibit a hydrodynamic instability leading to collective motion known as active turbulence, a phenomenon which is absent for pullers. In this Letter, we describe the collective dynamics of a binary pusher--puller mixture using kinetic theory and large-scale particle-resolved simulations. We derive and verify an instability criterion, showing that the critical density for active turbulence moves to higher values as the fraction $\chi$ of pullers is increased and disappears for $\chi \geq 0.5$. We then show analytically and numerically that the two-point hydrodynamic correlations of the 1:1 mixture are equal to those of a suspension of noninteracting swimmers. Strikingly, our numerical analysis furthermore shows that the full probability distribution of the fluid velocity fluctuations collapses onto the one of a noninteracting system at the same density, where swimmer--swimmer correlations are strictly absent. Our results thus indicate that the fluid velocity fluctuations in 1:1 pusher--puller mixtures are \emph{exactly} equal to those of the corresponding noninteracting suspension at any density, a surprising cancellation with no counterpart in equilibrium long-range interacting systems. 
\end{abstract}

\maketitle

A suspension of swimming microorganisms, such as bacteria or algae, is one of the archetypal examples of biological active matter at the microscopic scale~\cite{Cates1,Cisneros1,Lauga1,Marchetti1,Yeomans1}. At dilute concentrations, direct collisions between swimmers are rare, and interactions in biological microswimmer suspensions are therefore dominated by long-ranged hydrodynamic interactions leading to complex collective behaviour and significant swimmer-swimmer correlations~\cite{Baskaran1,Hohenegger1,Jepson1,Koch1,Lauga1,Qian1}. Arguably, the simplest description of biological microswimmers is that of a force dipole acting on the fluid, leading to a flow field that decays as the inverse square of the distance from the organism. In spite of its simplicity, the dipolar description of microswimmers has been shown to quantitatively describe the enhanced diffusion of passive tracer particles in \emph{E. coli} suspensions~\cite{Mino1,Jepson1,Morozov1}, as well as qualitatively explaining the onset of ``active turbulence'', whereby suspensions of bacteria undergo a transition to collective swimming characterised by significantly enhanced fluid velocities and long-ranged flow fields~\cite{Dunkel1,Gachelin1,Hohenegger1,Krishnamurthy1,Saintillan2,Sokolov1,Wensink1,Bardfalvy1}. Importantly, the transition to active turbulence as well as the build-up of pretransitional swimmer-swimmer correlations strongly depend on the sign of the force dipole~\cite{Stenhammar1,Qian1}, where active turbulence is only present for rear-actuated ``pusher'' microswimmers such as most bacteria. Their front-actuated counterpart, ``puller'' microswimmers, are less common in Nature: the bacterium \emph{Caulobacter crescentus} is able to switch between pusher and puller propulsion modes~\cite{Lele1}, and the front-actuated alga \emph{Chlamydomonas} oscillates between pusher and puller modes during its flagellar beat cycle~\cite{GuastoPRL,Klindt1}. While pure puller suspensions show no collective motion, models incorporating puller flow fields combined with short-ranged excluded volume interactions have been observed to induce a polar flocking state, both in pure puller suspensions~\cite{Alarcon:2013,Liverpool:PRE:2017,Menzel:JCP:2018} and in puller suspensions doped with a small pusher component~\cite{Menzel:MolPhys:2018}. This phase is driven by a combination of short-range collisions and mutual microswimmer reorientations due to long-ranged hydrodynamic interactions~\cite{Liverpool:PRE:2017,Menzel:JCP:2018}, and is fundamentally distinct from the nematically ordered active turbulent state characteristic of pusher suspensions~\cite{Dunkel1,Gachelin1,Hohenegger1,Krishnamurthy1,Saintillan2,Sokolov1,Wensink1,Bardfalvy1}.

\begin{figure}[h]
    \center
    \includegraphics[width=5cm,angle=-90]{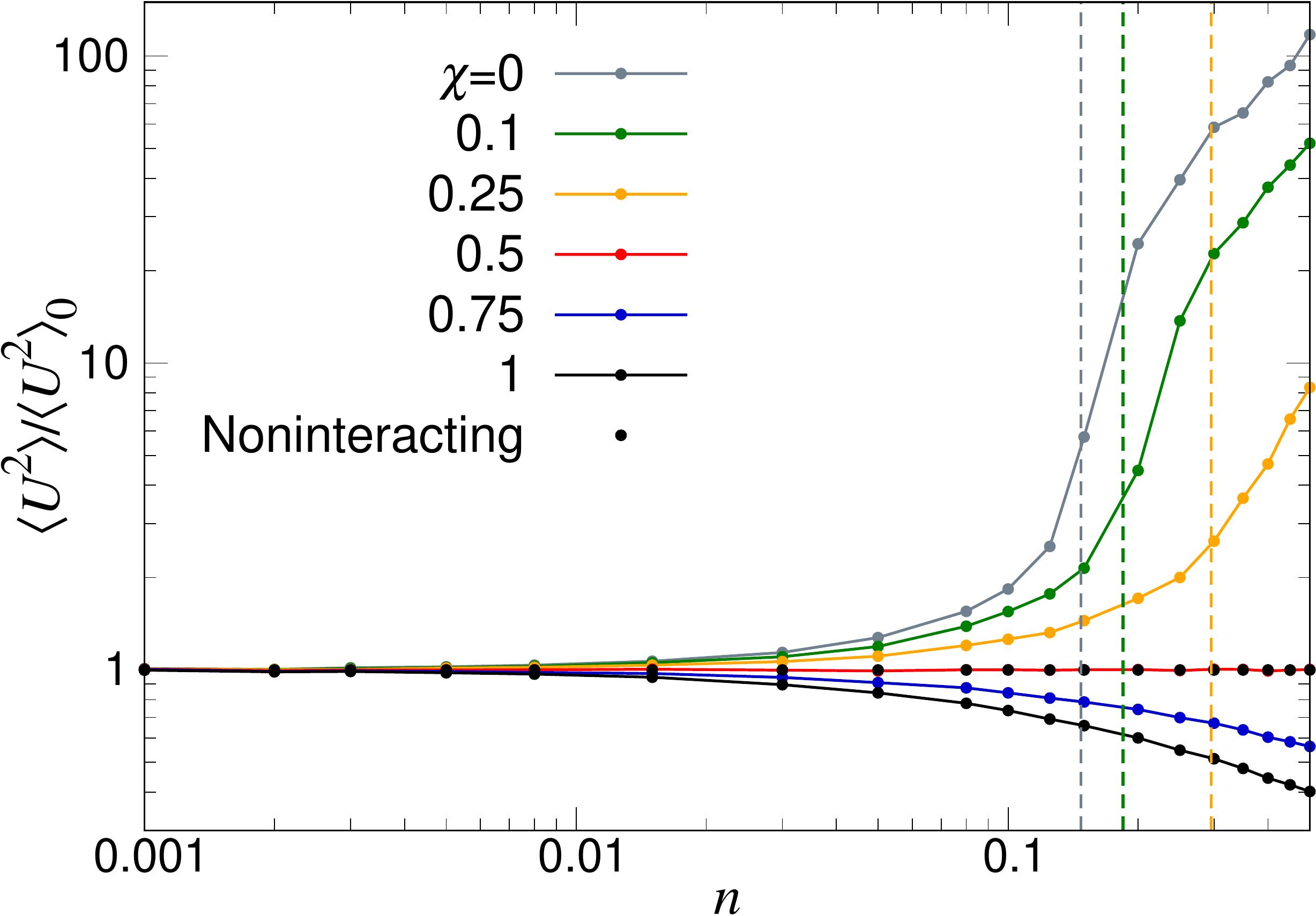}
    \caption{Fluid velocity variance $\langle U^2 \rangle$ as a function of the total number density $n$, normalized by the velocity variance $\langle U^2 \rangle _0$ of the corresponding noninteracting system for different puller fractions $\chi$. The dashed vertical lines show the predicted onset densities for active turbulence in pusher-puller mixtures according to Eq.~\eqref{eq:n_crit}, using the approximate value $n_c^{(0)} = 0.15$ for the threshold density in the pure pusher system, obtained by qualitatively taking into account the effect of periodic boundary conditions as described in~\cite{Bardfalvy1}.}\label{fig:1}
\end{figure}

In addition to its immediate biological relevance, the dipolar swimmer model has the advantage of being analytically tractable even at the many-body level. Thus, it constitutes an important minimal model of collective motion driven by long-ranged hydrodynamic interactions, in contrast to the short-range polar alignment interactions that induce collective motion in the Vicsek model~\cite{Vicsek1,chate2019dry} or direct steric collisions that cause motility-induced phase separation in active Brownian particles~\cite{Tailleur1,howse2007self,romanczuk2012active}.

At the continuum level, the transition to active turbulence in an infinite, unbounded suspension can be understood as a hydrodynamic instability occuring at a critical swimmer number density $n_c = 5\lambda / \kappa$~\cite{Saintillan1,Subramanian1,Stenhammar1,Saintillan4,Hohenegger1}, where $\lambda$ is the characteristic frequency of swimmer reorientation (tumbles), and $\kappa = Fl / \mu$ is the reduced dipolar strength with $F$ being the magnitude of the equal and opposite forces acting on the fluid, $l$ being their separation, and $\mu$ being the dynamic viscosity of the solvent. Crucially, this instability is purely driven by the mutual reorientations between pusher microswimmers, and is absent for pullers. No significant density inhomogeneieties have been observed in either pusher or puller suspensions regardless of the presence or absence of collective motion~\cite{Stenhammar1,Bardfalvy1}. While the instability itself can be inferred from a mean-field treatment, in order to capture the dynamics at intermediate microswimmer densities, but still below $n_c$, it is necessary to go beyond the mean-field description to include the effect of swimmer-swimmer correlations. Recent efforts~\cite{Stenhammar1,Skultety1,Qian1} have shown that correlations between microswimmers become significant at concentrations far below $n_c$. The mean-field description is thus only accurate for very dilute suspensions, where microswimmers can be described as effectively noninteracting, and pushers and pullers become statistically equivalent. 

In many ecosystems, bacteria and algae coexist, and the understanding of their mutual benefit or the parasitic behavior of one species on the other is currently considered a topic of great biological relevance~\cite{Ramanan1,Seymour1,Barbara1,Peaudecerf:PRE:2018}. Yet, at the collective level, very little is known about their mutual behaviour even in the simplified setting of the dipolar swimmer model, where bacteria and algae differ only through the signs of their force dipoles. Pessot \emph{et al.} have showed that the addition of a small amount of pushers suppresses the polar ordering emerging in 2-dimensional puller suspensions with excluded volume interactions due to the interplay between direct collisions and long-ranged hydrodynamics~\cite{Menzel:MolPhys:2018}. Moreover, Brotto \emph{et al.}~\cite{Brotto1} studied the related case of ``cyclic microswimmers'' that switch between pusher and puller modes within a mean-field framework, and found that the transition to collective motion vanishes when microswimmers spend more time in their puller state than in their pusher state. 

In this Letter, we reveal a set of novel, striking features of binary pusher--puller mixtures using particle-resolved lattice Boltzmann simulations and kinetic theory. We first show that the addition of pullers to a pusher suspension quickly increases the critical (total) density necessary for collective motion, which diverges for a 1:1 mixture. More strikingly, we find that the full spectrum of fluid velocity fluctuations in such symmetric mixtures \emph{exactly} overlap with those of a noninteracting microswimmer suspension where swimmer-swimmer correlations are strictly absent: In other words, the statistical properties of a 1:1 mixture are effectively those of an ``ideal gas'' of run-and-tumble microswimmers. Despite the presence of significant correlations among swimmers, this equivalence holds at \emph{any} density, a phenomenon for which we cannot find any analogy in equilibrium systems. 

\begin{figure}[h!]
    \center
    \includegraphics[width=7cm]{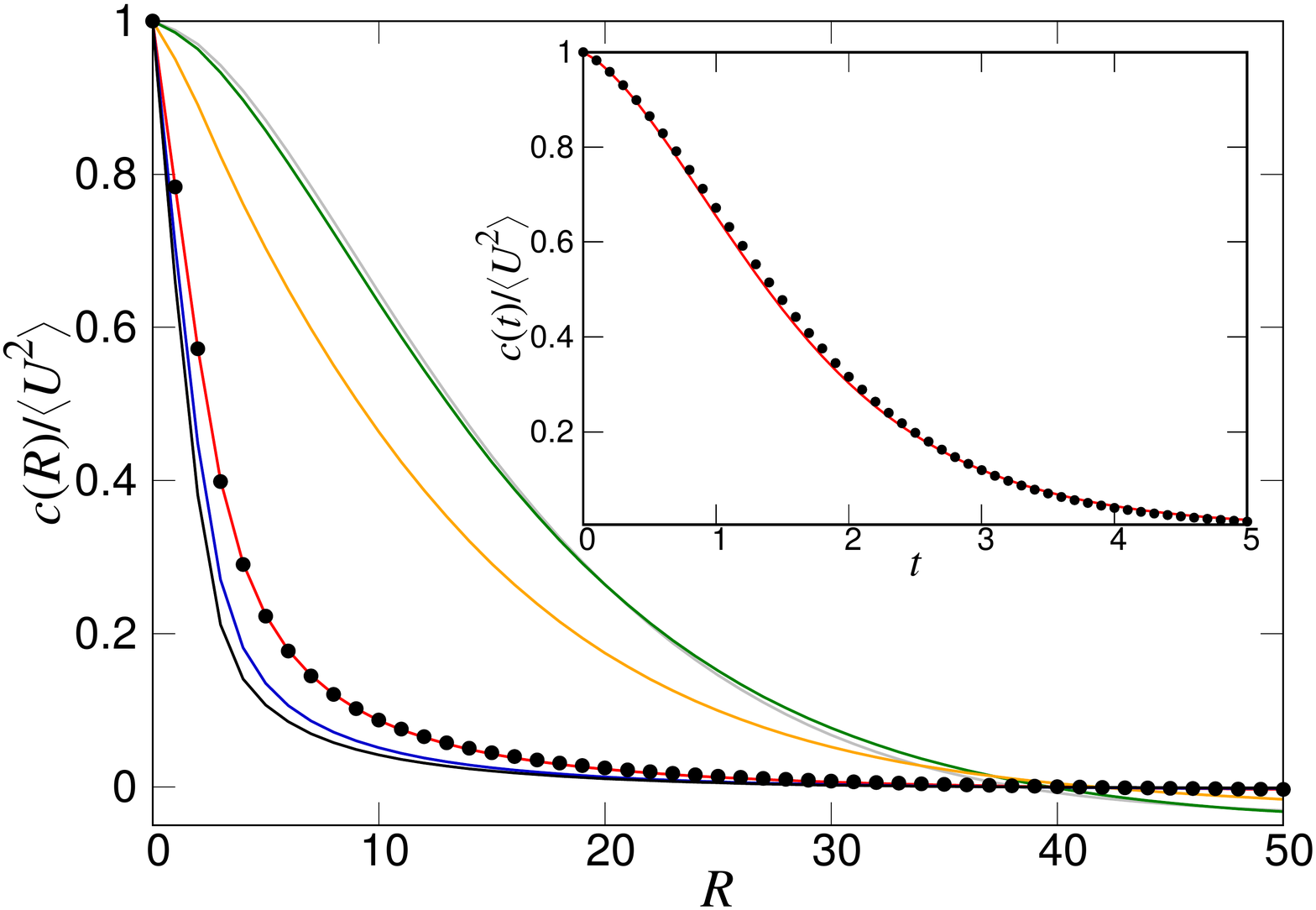}
    \caption{Normalised spatial correlation functions of the fluid velocity for $n=0.3$ and different $\chi$; color codes as in Fig.~\ref{fig:1}. Note that the results for $\chi = 0.5$ completely overlap with the corresponding ones for a noninteracting suspension. Inset: Temporal correlation function for $\chi = 0.5$ (line) and for a noninteracting suspension (symbols). }\label{fig:2}
\end{figure}

Our starting point is a 3-dimensional binary pusher-puller mixture of volume $V$ containing $N=N_{\mathrm{push}} + N_{\mathrm{pull}}$ microswimmers at total number density $n=N/V$; all densities considered here are in the dilute regime, where hydrodynamic interactions are dominant. Each swimmer is described by its dipolar strength $\kappa$, where we use the convention that $\kappa > 0$ corresponds to pushers and $\kappa < 0$ to pullers, and for simplicity assume that pusher and puller microswimmers only differ in their signs of $\kappa$. We furthermore define the puller fraction $\chi = N_{\mathrm{pull}}/N \in [0,1]$. The position $\mathbf{r}_i$ and orientation $\mathbf{p}_i$ of swimmer $i$ evolve according to the equations of motion 
\begin{align}
\dot{r}_i^\alpha &= v_s p_i^{\alpha}+U^\alpha(\mathbf{r}_i),\label{eq:rdot} \\
\dot{p}_i^\alpha &= (\delta^{\alpha\beta}- p_i^\alpha p_i^\beta) \nabla_i^\gamma U^\beta(\mathbf{r}_i) p_i^\gamma,\label{eq:pdot}
\end{align}
where $\delta^{\alpha\beta}$ is the Kronecker delta, $\mathbf{U}(\mathbf{r}_i)$ is the fluid velocity at the position of swimmer $i$, and $v_s$ is the (constant) swimming speed; Greek indices denote Cartesian components, and repeated upper indices are summed over. In addition to being rotated by the fluid, the swimmers' orientations are randomized with an average tumbling frequency $\lambda$. We solve the model using particle-resolved lattice Boltzmann (LB) simulations of up to $N = 5 \times 10^5$ microswimmers in a 3-dimensional periodic box of side $L=100$ in LB units; an in-depth description of the method can be found in~\cite{Bardfalvy1,Nash1}. Our parameters were chosen to approximately mimic those of an \emph{E. coli} suspension (see further~\cite{Bardfalvy1}): in LB units, we used $v_s = 10^{-3}$, $\lambda = 2 \times 10^{-4}$, $F = 1.55 \times 10^{-3}$, $l = 1$, and $\mu = 1/6$. In the following, we non-dimensionalise our results in terms of the swimmer length $l$ and the characteristic swimming time $l / v_s$. For simulations of noninteracting swimmers, we used a slightly higher swimming speed of $v_s = 1.03 \times 10^{-3}$ to compensate for the ``self-advection'' effect present in the interacting suspensions, leading to a slightly increased speed compared to the specified value; for a detailed discussion of this effect, see~\cite{Nash1}.

To understand the possible onset of collective motion in a binary pusher-puller mixture, we first consider the suspension from a mean-field kinetic theory perspective, along the lines of previous analyses for single-component suspensions and mixtures of cyclic microswimmers ~\cite{Saintillan1,Subramanian1,Stenhammar1,Saintillan4,Hohenegger1,Brotto1}.
We denote the one-body probability distribution functions for pushers and pullers by $f^+(\bfr,\bfp,t)$ and $f^-(\bfr,\bfp,t)$, respectively, with both functions normalized by $N$. We further consider the quantity $\Delta f = (1-\chi) f^+ - \chi f^-$, which can be used to express the fluid velocity within the mean-field approximation as 
\begin{equation}\label{eq:MF-U}
U^{\alpha}_{\mathrm{MF}}(\bfr,t)
=\int d\bfr_1 d\bfp_1 u_d^{\alpha}(\bfr-\bfr_1,\bfp_1) \Delta f(\bfr_1,\bfp_1,t),
\end{equation}
where $u_d^{\alpha}(\bfr-\bfr_1,\bfp_1)$ is the (regularised) dipolar flow field at $\bfr$ due to a swimmer at $\bfr_1$ with orientation $\bfp_1$~\cite{Stenhammar1}. At the mean-field level, $\Delta f$ evolves as
\begin{align}\label{eq:MF}
& \partial_t \Delta f + \lambda \Delta f- \frac{\lambda}{4\pi}\int d\mathbf{p} \Delta f  + \nabla^\alpha \big[ \left( v_s p^{\alpha}+U^\alpha_{\mathrm{MF}}\right) \Delta f \big]  \nonumber \\
& \qquad \qquad +  \mathbb{P}^{\alpha\beta}\frac{\partial}{\partial p^\beta} \big[ p^\gamma \mathbb{P}^{\alpha\delta} \left(\nabla^\gamma U^\delta_{\mathrm{MF}}  \right)\Delta f\big]
 = 0,
\end{align}
where $\mathbb{P}^{\alpha\beta} = \delta^{\alpha\beta} - p^\alpha p^\beta$. The linear stability of Eq.~\eqref{eq:MF} around the homogeneous and isotropic state can be obtained using standard methods developed for the single-component suspension~\cite{Saintillan1,Subramanian1,Stenhammar1,Saintillan4,Hohenegger1}. Similar to the single-component pusher case, the resulting instability sets in at the largest scale available to the system, at a critical density
\begin{equation}\label{eq:n_crit}
n^{(\chi)}_c = \frac{n^{(0)}_c }{1-2\chi},
\end{equation}
where $n^{(0)}_c = 5\lambda / \kappa$ is the critical density in an unbounded single-component pusher suspension~\footnote{The linear stability analysis presented in Eq.~\eqref{eq:n_crit} can be extended for multi-component mixtures with arbitrary properties. In the thin-rod limit, the instability criterion is given by $\sum_i n_i \kappa_i/\lambda_i = 5$, where $n_i$, $\kappa_i$ and $\lambda_i$ are respectively the number density, dipolar strength, and tumbling rate for species $i$, and the sum goes over all species in the mixture.}. Thus, active turbulence requires the concentration of pushers to be greater than that of pullers ($\chi < 0.5$), in analogy with the result previously derived for cyclic swimmers in~\cite{Brotto1}. In Fig.~\ref{fig:1}, we test this prediction by plotting the fluid velocity variance $\langle U^2 \rangle$, normalized by its value $\langle U^2 \rangle_0$ in a suspension of noninteracting swimmers where the terms containing $\bfU(\bfr)$ in the equations of motion \eqref{eq:rdot}--\eqref{eq:pdot} have been omitted. In accordance with previous findings for pure pusher suspensions~\cite{Stenhammar1,Bardfalvy1}, our simulations show that the expected sharp transition to collective motion is replaced by a rather smooth crossover (see Fig.~\ref{fig:1}). It is still not clear whether a sharp transition is recovered in the thermodynamic limit: simulation data with varying system sizes indicate that the smooth crossover is not a finite-size effect ~\cite{Stenhammar1,Bardfalvy1}, while recent theoretical results~\cite{Skultety1} suggest that strong swimmer-swimmer correlations below the transition  might change the sharp transition into a crossover~\cite{Skultety1}. Nevertheless, the position of the crossover for $\chi < 0.5$ is consistent with the prediction of Eq.~\eqref{eq:n_crit}, as shown by dashed vertical lines in Fig.~\ref{fig:1}. 

\begin{figure}[h]
    \center
    \includegraphics[angle=-90,width=7cm]{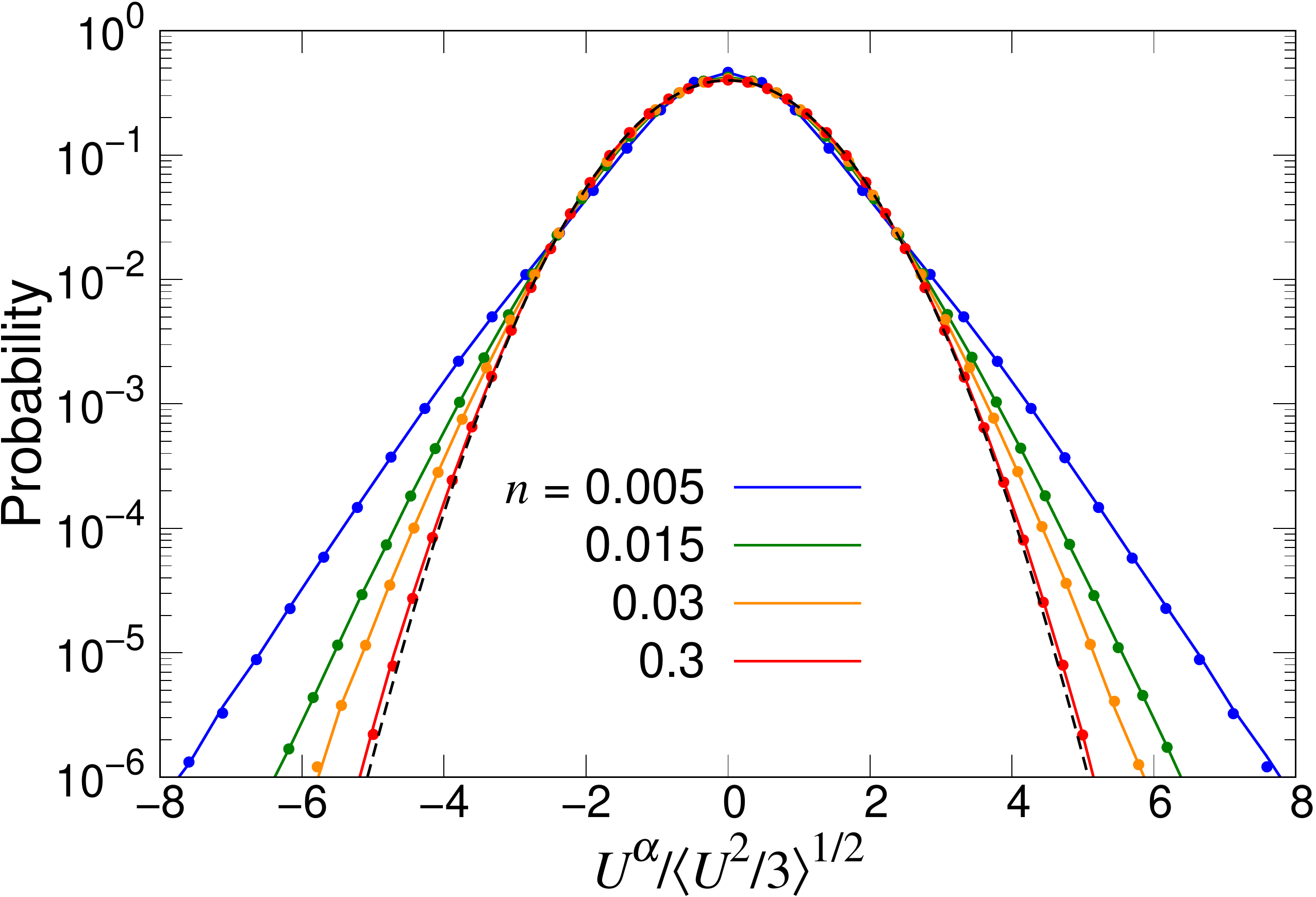}
    \caption{Normalized probability distribution function of the Cartesian components of the fluid velocity $\bfU$ at various densities, obtained from LB simulations. Solid lines show results for $\chi = 0.5$ and symbols for a noninteracting suspension at the same density. The black dashed line shows a unit-variance Gaussian function for comparison. The perfectly overlapping curves show that the velocity fluctuations in a 1:1 mixture, where swimmer-swimmer correlations are present, are equal to those in a noninteracting system, where they are strictly absent.}\label{fig:3}
\end{figure}

Strikingly, the 1:1 mixture ($\chi = 0.5$) exhibits a velocity variance that falls exactly on top of the corresponding value in a noninteracting suspension all the way up to the highest densities. This can be understood from a kinetic theory framework which has previously been developed for single-species systems in~\cite{Stenhammar1,Skultety1}; this theory can be generalised to mixtures, enabling the computation of any two-point observable. For the spatial correlation function $c(R)$ of the fluid velocity, we obtain
\begin{align}\label{eq:spatial-corr-1:1}
c(R) \equiv &\langle \bfU({\bf 0}) \cdot \bfU({\bf R})\rangle
=
\lim_{t\to\infty}
\int d\bfr_1\,d\bfr_2\,d\bfp_1\,d\bfp_2 
\\
&\langle h(\bfr_1,\bfp_1,t)h(\bfr_2,\bfp_2,t)\rangle
 u_d^\alpha(\bfr_1,\bfp_1) u_d^\alpha(\bfr_2-{\bf R},\bfp_2)\,
.\nonumber
\end{align}
where $h(\bfr,\bfp,t)$ are the phase-space fluctuations around the homogeneous and isotropic base state~\cite{Stenhammar1,Skultety1}. As in the single-species suspension, it can be shown that $h$ obeys the mean-field dynamics given by Eq.~\eqref{eq:MF} linearised around this base state and forced by a Gaussian noise that is independent of the interactions between swimmers. The analysis of the 1:1 mixture is hence particularly straightforward, as the dynamics of fluctuations is linearised around $\Delta f=0$:
\begin{equation}\label{eq:phase-space-fluct}
 \partial_t h +\lambda h -\frac{\lambda}{4\pi}\int d\bfp h = \xi
\end{equation}
where $\xi$ is a Gaussian process of zero mean and variance $\langle\xi(\bfr_1,\bfp_1,t_1)\xi(\bfr_2,\bfp_2,t_2)\rangle=2 \lambda (n / 4\pi) \left. \delta(t_1-t_2) \right. \left. \delta(\bfr_1-\bfr_2 ) \right. \left[ \delta(\bfp_1 - \bfp_2 ) - 1/(4\pi) \right]$. Since there is no dependence on the interactions in Eq.~\eqref{eq:phase-space-fluct}, it follows that any two-point observable that can be computed from the difference in phase-space densities of pushers and pullers equals the one in a noninteracting suspension at the same density. This analysis proves that, not only does the velocity variance of a 1:1 suspension coincide with the one for a suspension of noninteracting swimmers, but so does any observable that depends on two-point (spatial or temporal) correlations of the fluid velocity. In order to verify this claim, in Fig.~\ref{fig:2} we plot the spatial correlation function of the fluid velocity, defined by Eq.~\eqref{eq:spatial-corr-1:1}, in the concentration regime ($n=0.3$) where pure pusher suspensions show active turbulence. As expected, $c(R)$ decays faster as $\chi$ is increased, in accordance with the suppression of collective motion. Furthermore, the $\chi = 0.5$ data once more completely overlaps with the corresponding data for a noninteracting suspension all the way down to separations comparable to the swimmer size, verifying the equivalence in two-body correlations derived above. This equivalence is further confirmed by the overlap between the temporal correlation functions $c(t) \equiv \langle {\bf U}(0) \cdot {\bf U}(t)\rangle$ (Fig.~\ref{fig:2} inset). 

\begin{figure}[h]
    \center
    \includegraphics[angle=-90,width=7cm]{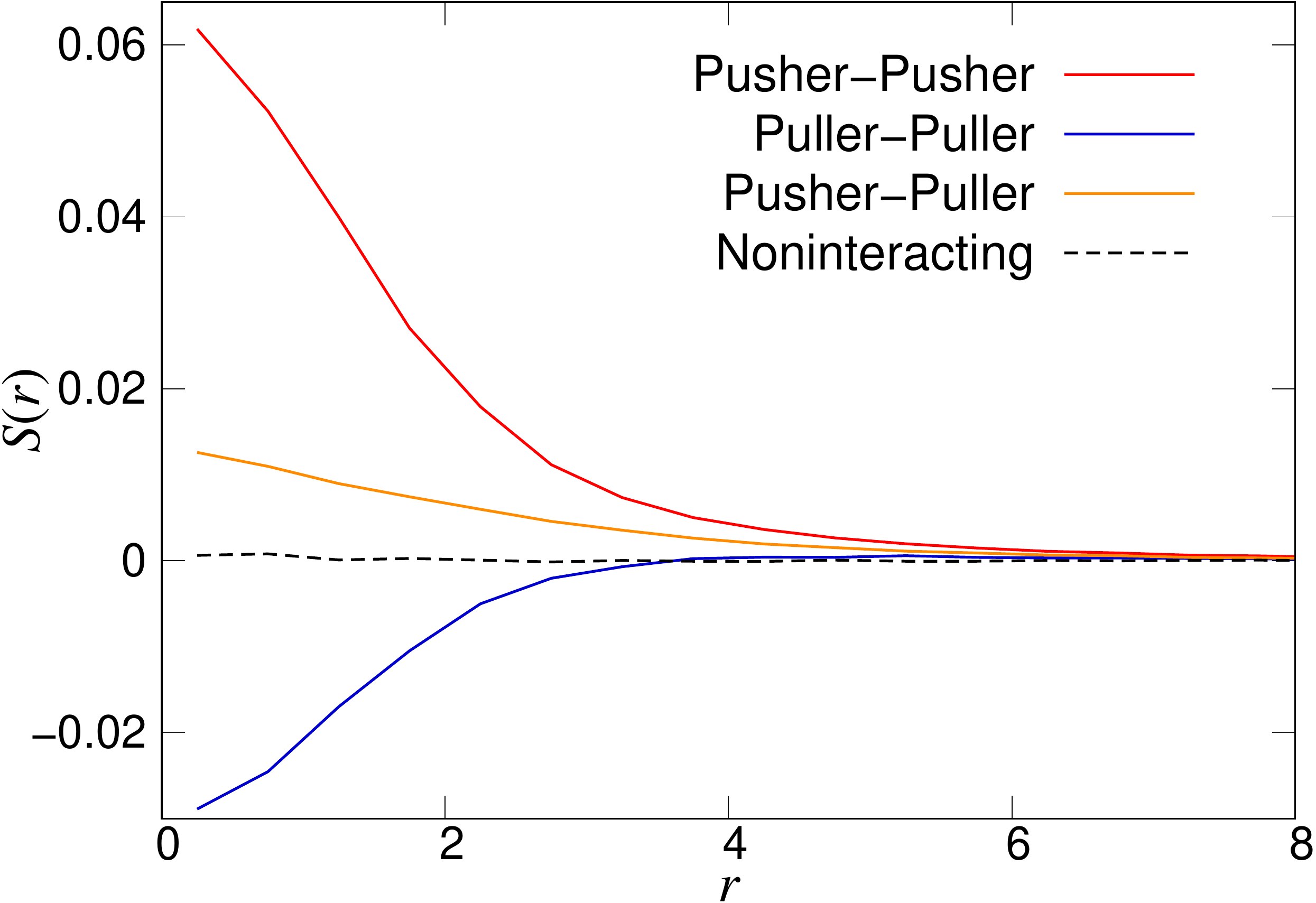}
    \caption{The three components of the pairwise swimmer-swimmer nematic order parameter $S(r)$ as a function of separation $r$ at $n=0.3$ and $\chi=0.5$. The dashed black line shows $S(r)$ in a noninteracting system where swimmer-swimmer correlations are strictly absent, and thus $S=0$. Note the significant local orientational correlations between swimmers in the $\chi = 0.5$ system, in spite of its velocity fluctuations being identical to those in the noninteracting system.}\label{fig:4}
\end{figure}

In order to investigate the striking equivalence between the 1:1 mixture and a noninteracting suspension further, we now go beyond the two-point correlations. In Fig.~\ref{fig:3}, we plot the full probability distribution (PDF) of the Cartesian components $U^\alpha$ for $\chi = 0.5$. At the lowest densities, the PDFs are strongly non-Gaussian, due to the relatively few swimmers contributing to the local fluid velocity in a single point~\cite{Zaid2011,Pushkin1}, while for the highest densities considered the PDF becomes Gaussian. Most strikingly, however, for all densities the PDFs overlap perfectly with the corresponding data for a noninteracting suspension. Our data thus suggests that there is an exact equivalence between the fluid velocity fluctuations in a 1:1 mixture and those in a noninteracting suspension, although we of course cannot exclude differences smaller than the statistical error or in the far tails of the distribution. Importantly, this cancellation between pusher and puller correlations is not due to the absence of orientational order between the swimmers. This is highlighted in Fig.~\ref{fig:4} where we show the separation-dependent nematic order between swimmers $S(r) = \langle P_2 (\cos \theta)\rangle_{|\mathbf{r}_i - \mathbf{r}_j| = r}$, with $P_2$ being the second Legendre polynomial, and $\theta$ the angle between $\bfp_i$ and $\bfp_j$; such nematic ordering between pusher swimmers is well-known to be associated with the transition to active turbulence~\cite{Bardfalvy1,Simha1}. Clearly, orientational swimmer-swimmer correlations are statistically significant, albeit weak and short-ranged, even in the 1:1 mixture. This is expected, as the swimmers are strongly interacting at the pairwise level, yielding a local nematic order between pusher swimmers and a weak antialignment ($S<0$) between pullers, in overall accordance with previous results in the low-density regime of single-species suspensions~\cite{Bardfalvy1}. 

The seemingly exact cancellation in the many-body dynamics of pusher and puller swimmers, leading to a distribution of fluid velocity fluctuations that exactly overlaps with the one of noninteracting swimmers, is highly nontrivial and surprising. We do not yet have an analytical understanding of this effect: computing higher moments than the second within the kinetic theory presented above would require taking into account nonlinear effects in the dynamics of phase-space fluctuations, and we see no obvious reason for them to give no contribution to the fluctuations of the fluid velocity. A possible method to investigate this phenomenon is large deviations theory, which has previously allowed progress in the study of rare fluctuations in minimal models of many-body interacting systems~\cite{derrida2007non,bertini2015macroscopic}. Nevertheless, analytical results are typically difficult to obtain and often achievable only perturbatively~\cite{bouchet2016perturbative}.

Moreover, to the best of our knowledge, the fact that fluctuations in the field that mediates interactions between particles are unaffected by the interactions themselves has no analogy in equilibrium long-range interacting systems: the simplest example of such a system is an electroneutral suspension of monovalent ions. There, however, even the two-point fluctuations in the potential and electric field depend strongly on the electrostatic coupling~\cite{Oosawa1}, and thus differ from the corresponding quantitites in the noninteracting limit. We conjecture that the cancellation observed here is rooted in the nature of effective action-reaction symmetry breaking in the dynamics of active matter systems, and would thus expect a similar cancellation to arise in other active matter systems such as mixtures of phoretic colloids, where fast diffusing chemicals induce Coulomb-like interactions that violate action-reaction symmetry~\cite{soto2014self}. An important direction for future work is thus to gain a deeper understanding of this phenomenon, which would constitute another important step forward in the study of the intriguing non-equilibrium collective dynamics of active matter systems.

\begin{acknowledgments} 
JS acknowledges funding from the Swedish Research Council (grant IDs 2015-05449 and 2019-03718). CN acknowledges the support of an Aide Investissements d'Avenir du LabEx PALM (ANR-10-LABX-0039-PALM). The simulations were performed on resources provided by the Swedish National Infrastructure for Computing (SNIC) at LUNARC.
\end{acknowledgments}

\bibliography{bibliography}
\end{document}